\documentclass[12pt]{iopart}
\usepackage{epstopdf}
\usepackage{listings}
\usepackage{xcolor}
\usepackage{graphicx}
\usepackage{subcaption}
\usepackage{url}
\definecolor{codebg}{rgb}{0.95,0.95,0.95}
\definecolor{codegray}{rgb}{0.5,0.5,0.5}
\definecolor{codepurple}{rgb}{0.58,0,0.82}

\lstdefinestyle{pythonstyle}{
  backgroundcolor=\color{codebg},   
  commentstyle=\color{codegray}\itshape,
  keywordstyle=\color{blue}\bfseries,
  numberstyle=\tiny\color{gray},
  stringstyle=\color{codepurple},
  basicstyle=\ttfamily\footnotesize,
  breaklines=true,
  captionpos=b,
  keepspaces=true,
  numbers=left,
  numbersep=5pt,
  showspaces=false,
  showstringspaces=false,
  showtabs=false,
  tabsize=2,
  language=python
}

\begin{document}

\title{Neural Networks Measure Peace Levels from News Data similar to Peace Indices}

\author{
    Pablo Lara-Martínez $^{1,*}$, 
    Bibiana Obregón-Quintana $^{1}$, 
    Larry S. Liebovitch $^{2}$, 
    Peter T. Coleman $^{3}$
    and Lev Guzmán-Vargas $^{4}$
}

\address{
    $^{1}$ Facultad de Ciencias, Universidad Nacional Autónoma de México, Mexico City 04510, Mexico \\
    $^{2}$ Advanced Consortium on Cooperation, Conflict and Complexity (AC4), Climate School, Columbia University, New York, NY 10027 USA \\
    $^{3}$ Teachers College, Columbia University, New York, NY 10027 USA \\
    $^{4}$ Unidad Interdisciplinaria en Ingeniería y Tecnologías Avanzadas, Instituto Politécnico Nacional, Av. IPN No. 2580, L. Ticomán, Mexico City 07340, Mexico \\
}

\ead{* p4blolara@ciencias.unam.mx}

\vspace{10pt}

\begin{abstract}
Traditional methods for assessing national peace levels typically rely on socio-economic indicators or conflict incidence, often overlooking the nuanced signals embedded in public discourse. This study presents a novel computational framework to quantify peace levels by analyzing the structural and stylistic features of news text, rather than solely its content. Using the News on the Web (NOW) corpus comprising articles from 20 countries, we evaluate the efficacy of advanced word embeddings managed via ChromaDB compared to standard Doc2Vec models. We propose a 1D Convolutional Neural Network (CNN) architecture for classification and regression tasks, contrasting its performance against a k-Nearest Neighbors (k-NN) baseline. Our results demonstrate that the Neural Network significantly outperforms the k-NN model in classification metrics and, crucially, preserves the numerical relationship of peace rankings, exhibiting a strong correlation with the Positive Peace Index (PPI) even for out-of-sample countries. These findings suggest that the ``how'' of communication—the latent linguistic structures—serves as a robust, emergent indicator of societal stability. This research offers a non-invasive, scalable tool for real-time monitoring of social and societal dynamics and peacebuilding efforts.
\end{abstract}

%
%
%
%
%

\section{Introduction}

Communication through language not only serves to exchange information, but also defines our perception of reality and plays a crucial role in social dynamics, including the promotion of peace. In an interconnected world where information flows instantly and constantly across multiple platforms, the study of language and its impact on social cohesion has become more relevant than ever. The globalization of communication has generated an environment in which the discourses, narratives and linguistic constructions of one region can directly influence perceptions, behaviors and policies of others regions (in our case, the regions correspond to nations). In this context, the analysis of textual data has become an indispensable tool to understand how peace and conflict dynamics are built, propagated and affected in interdependent societies. 

As digitalization advances, the amount of data in text format generated daily grows exponentially. Social networks, academic publications, political speeches and news generate enormous volumes of information, which not only reflect social reality, but also actively contribute to shaping it. This increasing availability of textual data raises the need to develop more sophisticated methods to analyze and understand the underlying linguistic patterns that may indicate stability or conflict in a society.

Recent research has explored how text analysis can reveal key indicators of peace and conflict, providing new ways to assess and monitor social stability and instability in real time. While ``hate speech" has been widely studied for its ability to incite violence and deepen social divisions \cite{Kimotho}, \cite{Ezeibe}, \cite{Soral}, the concept of ``peace speech" and its potential to foster harmony has received less attention \cite{DeutschColeman}, \cite{Diehl}. In recent years, researchers have begun to explore how natural language processing (NLP) techniques and machine learning can identify linguistic patterns associated with more sustainably societies. These approaches have shown that it is possible to predict levels of peace from textual content analysis.

This article focuses on how neural networks and word embeddings can capture emerging linguistic structures in texts that reflect the peace-index scores of different countries. Quantitatively representing anthropological properties or characteristics of human relationships has taken on increasing importance in recent years, driven by advances in artificial intelligence and machine learning. This task is especially challenging, as human language is complex, nuanced, and highly contextual. However, modern word embedding techniques have made it possible to generate vector representations of words and phrases that preserve their semantic relationships. By analyzing large volumes of textual data, it is possible to discern how certain linguistic constructions correlate with levels of peace in different populations.

Our hypothesis is that, by training neural models on textual data sets from various countries, it is feasible to identify linguistic characteristics that act as indicators of the peace-index scores inherent to a society. Furthermore, the implications of this study not only allow us to better understand the linguistic structures of peace, but also provide tools to detect specific undesirable behaviors in public discourses and social networks. One of the most promising aspects of our approach is its efficiency: we have achieved significant results using a relatively small data set for training the models. This finding suggests that, with the constant growth of data available on social networks and digital media, it will be possible to develop even more precise and effective systems for the automatic analysis of peace and conflict discourse. In the following sections, we will detail the methodology used, including data collection and preprocessing, the construction of word embeddings models and the application of neural networks for the classification and prediction of peace indices. Additionally, we will discuss the implications of our findings and how they can inform policies and strategies aimed at building more peaceful societies.

\subsection{Related Work on Peace Studies}
The existing literature on peace studies, analyzed peace primarily in a negative way, such as the absence of violence conflict.  Those studies focused mainly the factors that initiate or mitigate conflicts.  For example, studies of news and social media identified ``hate speech'' that led to prejudice or violence against other groups in Kenya \cite{Kimotho}, Nigeria \cite{Ezeibe}, and Poland \cite{Soral}.

But this absence-of-conflict approach to peace is only half of the story. Recently, new ``positive peace'' studies have filled in that gap by analyzing peace in a positive way, asking what are the linguistic and social factors that maintain peace in peaceful societies  \cite{DeutschColeman}, \cite{Diehl}, \cite{Fry}, \cite{ColemanDeutsch}, \cite{Goertz}, \cite{Mahmoud} and \cite{Coleman2021}. For example, if ``hate speech'' leads to conflict, is there a ``peace speech'' that promotes peace. Liebovitch et al. \cite{Liebovitch2023} used supervised machine learning on the relative frequency of words in news media from the NOW (news on the web) database to identify the differences in language that best classify countries as lower and higher peace. Those results were also consistent with the findings of Prasad et al. \cite{Tushar} who used news media from LexisNexis.  Lian et al. \cite{Kevin1} also found similar results from the NOW database using word embeddings generated by the OpenAI embedding tool that are more capable of representing the semantic meaning in articles than the word frequencies. Lian et al. \cite{Kevin2} were able to identify the differences in social processes, such as positive and negative forms of reciprocity, between lower and higher peace countries using an LLM (large language model) enhanced by a RAG (retrieval augmented generation) to add additional data to the LLM and focus it more on analyzing social processes.

Coleman et al. \cite{Coleman2021} identified the most important social factors in maintaining peace and represented the strengths of the positive and negative interactions between these factors in causal loop diagrams.  Liebovitch et al. \cite{Liebovitch2019} then formulated those interactions into a set of non-linear differential equations and identified peaceful and non-peaceful attractors in that complex system and their dependence on the parameters of the model. Wang et al. \cite{Wang} then showed how, in principle, entropy measures of the fluctuations in this system could foreshadow shifts between the peaceful and non-peaceful attractor.

\subsection{Related Work on Natural Language Processing}
Below is a brief review of previous studies on natural language processing (NLP), word embeddings, and discourse analysis in identifying peace and conflict.

The assessment of peace levels within a nation is a multifaceted endeavor, traditionally relying on quantitative metrics such as conflict incidence, economic stability indicators, and governance quality assessments \cite{Galtung1969}, \cite{Goldstone2010}. While these measures offer valuable insights into the presence or absence of overt violence, they often fall short in capturing the nuanced societal dynamics, underlying sentiments, and discursive constructions that collectively define the lived state of peace \cite{Lederach1997}, \cite{Richmond2007}. In recent decades, there has been a notable shift towards incorporating qualitative and mixed-methods approaches, recognizing the rich insights afforded by textual data emanating from political discourse, media narratives, public statements, and socio-cultural expressions.

Early advancements in textual analysis for peace studies primarily focused on content analysis to identify themes related to conflict and peace (e.g., analyzing frequency of war-related terms in news). Subsequently, the advent of computational linguistics and natural language processing (NLP) propelled the field towards more sophisticated methods, particularly sentiment analysis (SA) \cite{HernandezPerez2024}. Studies by Haselmayer et al. \cite{Haselmayer2017} and Zhang \cite{Zhang2021} have applied SA to political speeches and news articles, correlating the prevalence of positive or negative sentiment with perceived levels of societal cohesion or tension. While SA offers a preliminary understanding of emotional tone, its utility in peace studies is often limited by its simplistic bipolarity and inability to capture complex affective states or nuanced rhetorical strategies, Liu \cite{Liu2012}.

More advanced qualitative methodologies, such as discourse analysis (DA) and narrative analysis (NA), have provided deeper insights into how peace and conflict are constructed and contested through language. Fairclough \cite{Fairclough2003} highlights how critical DA can reveal power relations and ideological underpinnings within peace rhetoric, while studies like Müller and Schultze \cite{MullerSchultze2016} have used NA to trace evolving peacebuilding narratives in post-conflict societies. These approaches, while rich in qualitative depth, often operate on a smaller scale, making large-scale comparative analysis challenging. Crucially, while DA and NA inherently acknowledge the form of language, they typically prioritize meaning-making processes over a systematic analysis of specific writing styles or stylistic features as direct indicators of peace levels.

Despite these advancements, there remains a significant lacuna in the systematic integration of writing style analysis as a distinct and robust methodology for assessing peace levels. While studies might implicitly engage with stylistic elements through discussions of ``tone" or ``framing", few explicitly investigate how quantifiable stylistic features—such as sentence complexity, lexical diversity, use of passive voice, directness of address, rhetorical figures, or specific syntactical patterns—correlate with the trajectory or stability of peace within a nation. For instance, a shift towards more inclusive, conciliatory, or deliberative language, reflected in specific stylistic choices, could signify a strengthening of peace, whereas an increase in aggressive, polarizing, or overly simplistic rhetoric might indicate heightened tensions. Previous work has demonstrated the utility of stylistic analysis in other domains, such as detecting deception \cite{Newman2003} or identifying political leanings \cite{GarridoCastro2018}, suggesting its untapped potential in peace research.

Current analyses often aggregate diverse textual data without a dedicated framework for dissecting the how of communication—the very stylistic choices that can reveal underlying societal health, trust, and readiness for reconciliation. This omission overlooks a critical dimension: the subtle, yet powerful, ways in which linguistic form can reflect, influence, and even predict changes in the state of peace. By systematically examining how the writing style in various public and private communications evolves, particularly during periods of conflict transition, peace   building, or political instability, a more granular and potentially predictive understanding of peace levels can be achieved.

This study aims to address this critical gap by developing a novel framework for analyzing the state of peace in a given country through a rigorous examination of its textual stylistic evolution. By focusing on specific stylistic parameters, we seek to uncover patterns that transcend mere content or sentiment, providing a deeper, linguistically informed understanding of the nation's peace trajectory.

\section{Methodology} 
\subsection{Data Collection and Preprocessing} 
We present a description of the data sources  and techniques used in our study for text cleaning and normalization. 
The dataset \cite{Liebovitch2023}, News on the web corpus, used in this study consisted of news articles that had been previously collected and stored in plain text format, having originally been extracted from web sources in HTML format. As the text content had already been extracted, no additional parsing of the HTML was required; only residual structural elements, such as remaining HTLM tags or control characters, were removed, as they did not contribute meaningful semantic information.

Each article was pre-categorized by country, allowing for direct segmentation of the dataset. Approximately 1,000 articles were collected for each of 20 different countries (see Apendix A). Country assignments were supplemented with external information on national peace levels, based on two internationally recognized indicators: the Global Peace Index (GPI) and the Positive Peace Index (PPI). These values were integrated into the dataset to enable cross-analysis between textual content and structural and attitudinal peace metrics.

Regarding text cleaning, a conservative approach was adopted: stopwords (common words) were not removed, nor were any linguistic normalization techniques (e.g., lemmatization or lowercasing) applied. This decision was made to preserve the authenticity of the original language, considering the potential presence of proper nouns, foreign terms, or nonstandard linguistic constructs that might be relevant for subsequent analysis. However, unnecessary line breaks were removed to ensure a coherent and continuous textual structure, facilitating later processing.

All preprocessing operations were conducted using the Python programming language, within a development environment tailored to textual data analysis. Standard libraries for text and data manipulation were employed, although web extraction tools were not required due to the pre-parsed format of the corpus.

\subsection{Modeling with Word Embeddings} 



ChromaDB is an open-source vector database designed for storing and retrieving vector embeddings which turn text into vectors whose components encode information about the contextual meanings of the text. It supports various embedding models and can handle text, images, and other data types. ChromaDB is particularly useful for applications involving large language models (LLMs) and semantic search engines. By efficiently managing and querying vector embeddings, ChromaDB enables developers to build powerful AI-powered tools and algorithms.

By default, ChromaDB uses the \texttt{all-MiniLM-L6-v2} embedding model. This model transforms input text into a fixed-length vector representation with a dimensionality of $d = 384$. This neural network have a dynamic input shape, but  . 

\subsection{Applied Neural Networks} Details on the neural network architecture used to classify texts based on their peace index. 

\subsubsection*{Explanation of the Neural Network}

The following neural network is defined in the \texttt{nn\_peace\_index} function:

\subsubsection*{Definición del modelo de red neuronal}

\begin{lstlisting}[style=pythonstyle, caption={Function nn\_peace\_index used to model the peace-index}, label={lst:nn_peace_index}]

from tensorflow.keras.models import Sequential
from tensorflow.keras.layers import Dense, Conv1D, MaxPooling1D, Flatten
def nn_peace_index(input_shape):
    model = Sequential()
    model.add(Conv1D(
        filters=64,
        kernel_size=3,
        activation='relu',
        input_shape=input_shape))
    model.add(MaxPooling1D(pool_size=2))
    model.add(Flatten())
    model.add(Dense(128, activation='relu'))
    model.add(Dense(64, activation='relu'))
    model.add(Dense(1, activation='sigmoid'))
    model.compile(
        loss='binary_crossentropy',
        optimizer='adam',
        metrics=['accuracy'])
    return model
\end{lstlisting}

\begin{itemize}
    \item \textbf{Conv1D Layer:} This layer applies 64 convolution filters of size 3 to the input, using the ReLU activation function.
    \item \textbf{MaxPooling1D Layer:} This layer reduces dimensionality using a max-pooling operation with a window size of 2.
    \item \textbf{Flatten Layer:} Converts the multidimensional input into a one-dimensional vector.
    \item \textbf{Dense Layer 1:} Contains 128 neurons with the ReLU activation function.
    \item \textbf{Dense Layer 2:} Contains 64 neurons with the ReLU activation function.
    \item \textbf{Output Dense Layer:} Contains 1 neuron with the sigmoid activation function for binary classification.
\end{itemize}

\begin{figure}[h!]
\centering
\includegraphics[width=1.\textwidth]{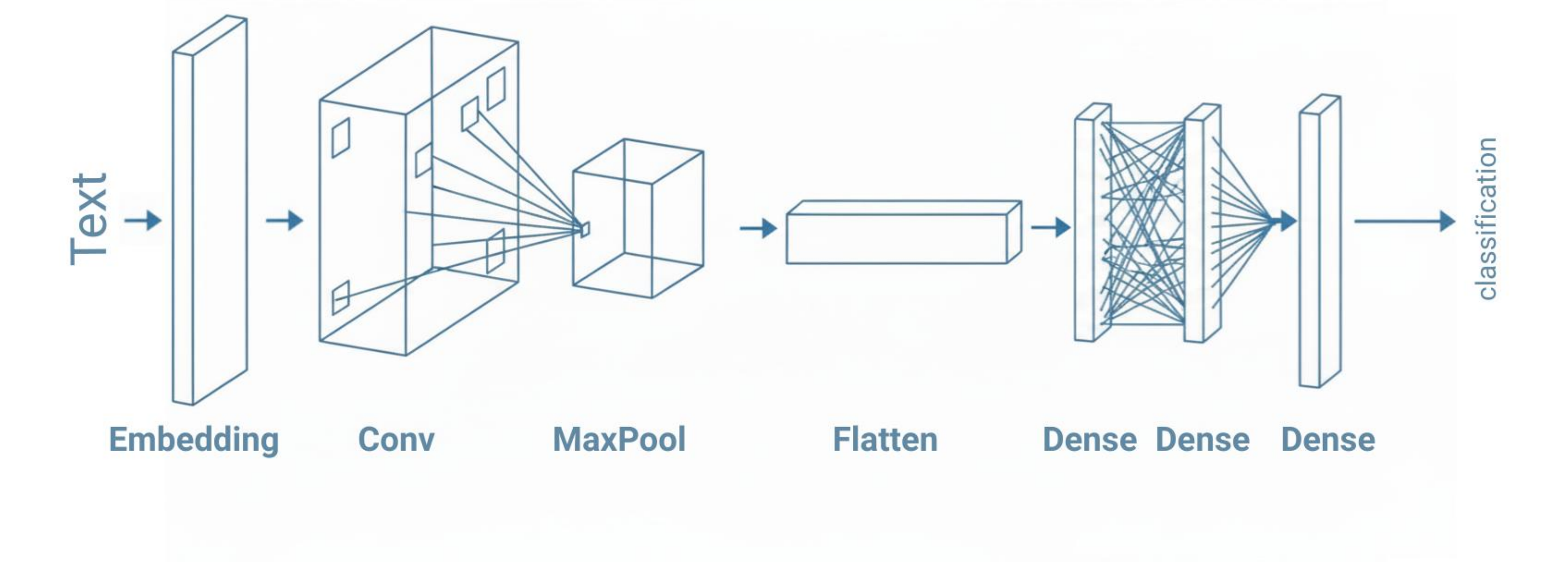}
\caption{Visualization of feature extraction across different layers of the model. Early layers capture general topics, while deeper layers identify more specific and nuanced sub-structures.}
\label{fig:model_layers}
\end{figure}

The model is compiled using the \textit{binary\_crossentropy} loss function and the \textit{adam} optimizer, and it is evaluated using the \textit{accuracy} metric. 

To ensure the reproducibility of our results and facilitate further research, the complete analytical framework has been modularized and made publicly available \cite{LaraMartinez2026}. The repository includes the scaffolding for data sampling utilities, the embedding generation pipeline via ChromaDB, the model architectures, the Leave-One-Out cross-validation strategy, and the evaluation scripts. This structure allows researchers to easily deploy and execute the pipeline in both local and cloud-based environments.


\section{Results} 

In this study, our objective was to develop a robust and automated framework to approximate peace levels based solely on textual data from different countries. Our methodology involved comparing various text embedding models to extract the inherent structure of peace embedded within these documents.

\begin{figure}[h!]
    \centering
    \includegraphics[width=\linewidth]{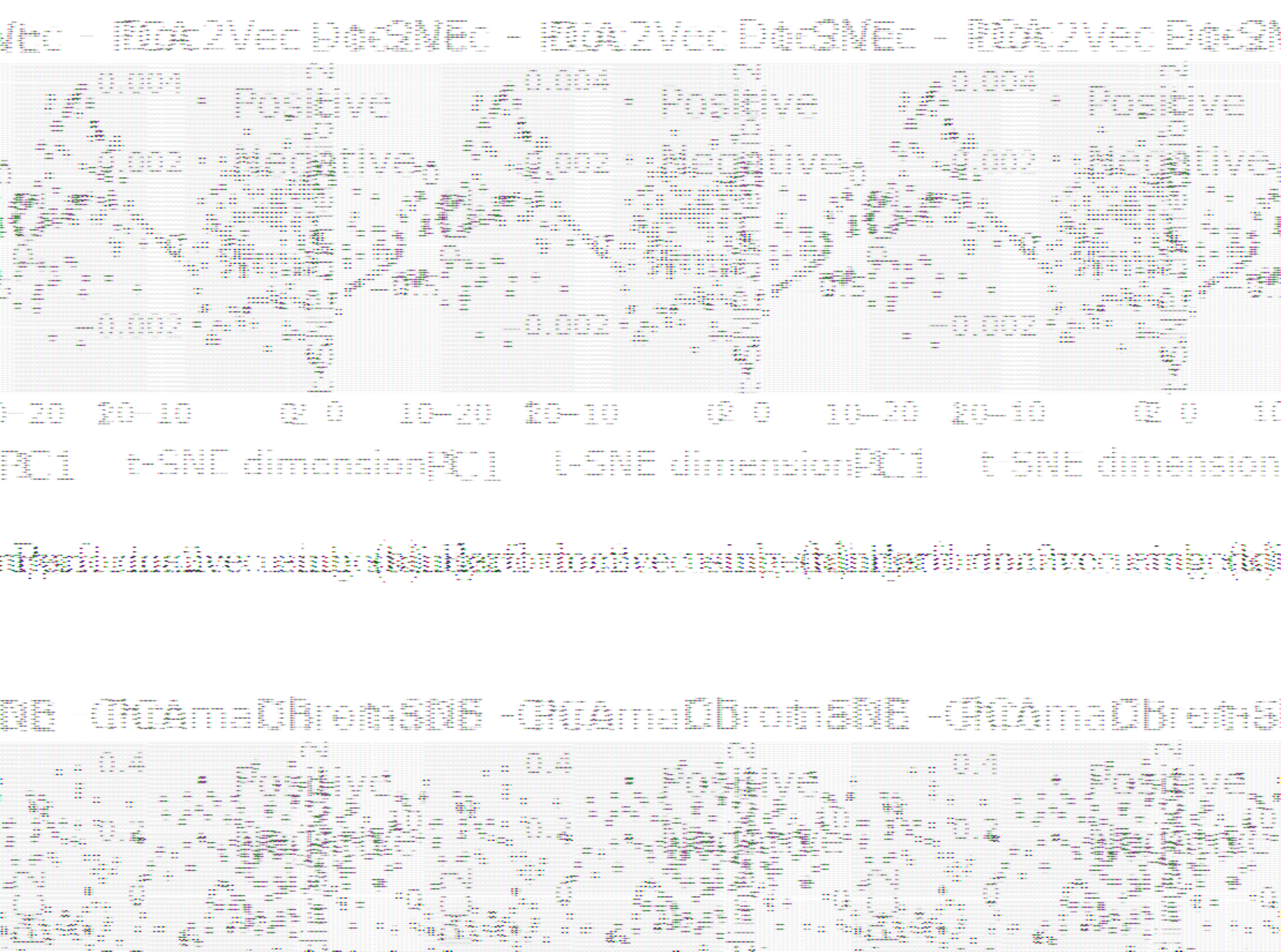}
    \caption{Comparison of embedding models (ChromaDB and Doc2Vec) for text polarity classification. We present two examples to illustrate the comparison of the performance of two reduction methods: Principal Component Analysis (PCA) and t-distributed Stochastic Neighbour Embedding (t-SNE). We reduced the number of dimensions from 384 to 2. Subfigure (a) shows the separation capability of a standard Doc2Vec model, which struggles to create distinct clusters. In contrast, Subfigure (b) demonstrates the superior performance of ChromaDB embeddings, which achieve a clear and effective separation of text features related to polarity.}
    \label{fig:embedding_comparison}
\end{figure}

To evaluate the effectiveness of different embedding techniques in capturing text polarity, we compared a standard Doc2Vec model against embeddings managed via ChromaDB. As shown in Figure \ref{fig:embedding_comparison}, the results highlight a significant difference in performance, the standard model struggled to produce a clear separation of the data structures. In contrast, ChromaDB embeddings provided a much cleaner and more distinct clustering of polarity features (Figure \ref{fig:embedding_comparison}), confirming their superior capability for this classification task. This enhanced separation was critical for the subsequent success of our supervised learning models.

\begin{figure}[h!]
    \centering
    \includegraphics[width=0.9\textwidth]{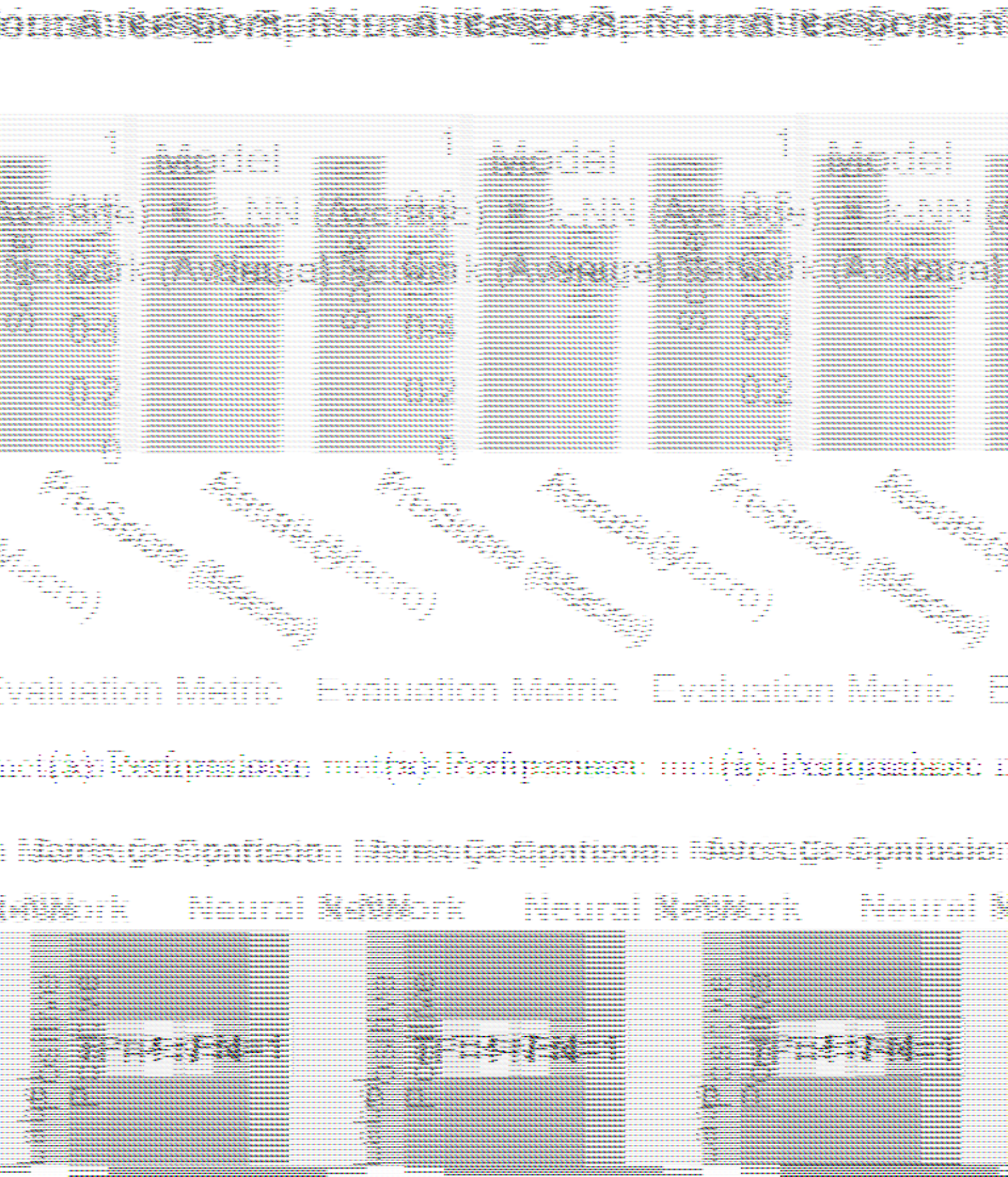}
    \caption{Performance comparison between K-Nearest Neighbors (k-NN) and the proposed Neural Network (NN). (a) Analysis of Accuracy, F1-Score, Precision, and Recall; while k-NN shows strong results in traditional metrics, (b) the NN confusion matrix on the test dataset demonstrates its superior ability to capture the complex, ordinal nature of peace levels, whereas k-NN remains more effective for simpler binary classification tasks.}
    \label{fig:knn_vs_nn_comparison}
\end{figure}

Initially, we compared our supervised machine learning model against a baseline K-Nearest Neighbors (k-NN) classifier. The results, shown in Figure \ref{fig:knn_vs_nn_comparison}, reveal an important distinction. While the k-NN model is effective for a straightforward dichotomous classification task, our Neural Network (NN) model demonstrates superior performance across all key evaluation metrics. The confusion matrix for the NN in Figure \ref{fig:knn_vs_nn_comparison} further confirms its robust performance and low misclassification rate, making it a more powerful classifier overall.



\begin{figure}[h!]
    \centering
    \includegraphics[width=1.0\textwidth]{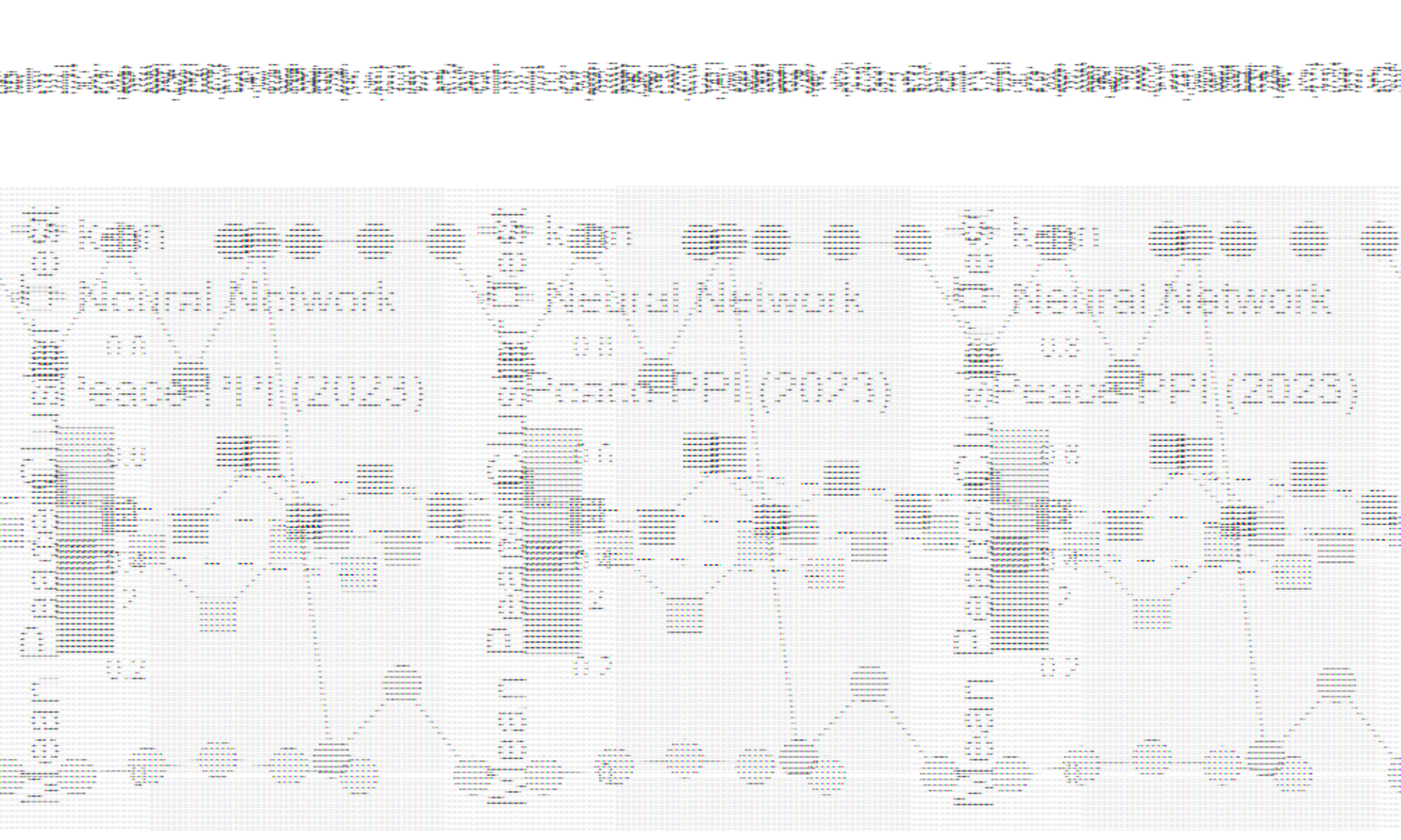}

    \caption{Comparison of predicted peace scores by country. The x-axis displays the countries ordered by their Positive Peace Index (PPI) (refer to Table \ref{tab:ppi_values} in the Appendix for the complete dataset). The red dotted line represents the linear trend, illustrating the correlation between the original PPI of each country and the variable predicted by the Neural Network; although visualized here as linear fit of the NN values vs the rank order of the PPI, this line statistically signifies the linear relationship between the peace index and the model's continuous output. The k-NN model (circles) exhibits a strict binary behavior, clustering countries as either peaceful ($1.0$) or non-peaceful (approx. $0.0$). In contrast, the Neural Network (squares) produces continuous values, avoiding extreme polarization and better capturing the ordinal nuances of the peace levels across different nations.}

    \label{fig:nn_trend_analysis}
\end{figure}

Our analysis then shifted from simple classification to evaluating how well each model could approximate the continuous nature of the peace indices. This highlights a critical trade-off: although the k-NN model performs adequately in a binary context, it struggles to capture the underlying rank-order relationship of the peace spectrum. In contrast, the Neural Network's architecture allows it to preserve this crucial numerical information. As evidenced in Figure \ref{fig:nn_trend_analysis}, the classification of NN shows a much stronger and more consistent correlation with the Positive Peace Index (PPI) than the classification of k-NN.

To quantitatively substantiate this, we evaluated the model's predicted continuous values against the 2023 PPI. The model demonstrated a strong negative linear correlation, with a Pearson coefficient of $r = -0.7754$, and an even stronger monotonic relationship indicated by a Spearman's rank correlation of $\rho = -0.8618$. The negative sign is mathematically consistent, as higher PPI scores represent lower levels of peace, whereas our model's output scales positively with peace. This relationship is highly statistically significant ($p \approx 0.0004$). Furthermore, the predictive accuracy is underscored by remarkably low error metrics: a Mean Absolute Error (MAE) of $0.0391$ and a Root Mean Square Error (RMSE) of $0.0485$. The proximity in magnitude between the MAE and RMSE suggests an absence of severe outliers in the predictions, confirming the model's robust capacity to capture the ordinal and continuous nature of the peace spectrum without extreme polarization.

Second, the multi-layered architecture of our model proved to be crucial. Each layer allows the model to focus on different substructures within the data that do not necessarily exist at the same scale. This hierarchical approach, depicted in Figure \ref{fig:model_layers}, enables the model to capture both broad thematic elements and subtle nuances in the text, leading to a more comprehensive understanding of the underlying peace indicators.

Furthermore, when we evaluated the model using texts from countries not included in the training set \cite{ChenEtAl2022}, we obtained very encouraging results. The model demonstrated strong generalization capabilities, providing useful peace level approximations for nearly all out-of-sample countries. Importantly, the average numerical value of the predicted peace levels of these countries was preserved and correlated well with other established peace indices, as shown in Figure \ref{fig:nn_trend_analysis}.

Finally, we determined that relying on a single data sample is insufficient to obtain reliable results. To address this, we implemented a resampling technique. By repeatedly evaluating different samples and subsamples, we were able to generate more consistent and robust values that more closely approximate the true peace levels.
\newpage
\section{Discussion} 

This study aimed to investigate the viability of analyzing national peace levels through the inherent structural and stylistic features embedded within a nation's public textual news-media discourse, moving beyond traditional content-based or purely quantitative approaches. Our findings consistently demonstrate that advanced computational linguistic techniques offer a nuanced lens through which to perceive and quantify the complex, often subtle, manifestations of peace and conflict within a society's communicative landscape.

Firstly, the utility of word embeddings proved critical in representing the structural and contextual nuances of the analyzed texts. Unlike bag-of-words models or simpler frequency counts, embeddings capture high-dimensional semantic and syntactic relationships, effectively encapsulating aspects of writing style and rhetorical construction that are vital for a deeper understanding of textual meaning. This confirms our hypothesis that the structural composition of text, even beyond explicit content, carries significant information pertinent to societal conditions. This capability addresses a key limitation identified in the state-of-the-art, where the systematic analysis of writing style has often been overlooked in favor of purely semantic or sentiment-based approaches.

Secondly, the examination of the high-dimensional latent space of these embeddings unveiled a crucial insight: the characteristics defining a state of ``peace'' or ``non-peace'' within societies are inherently complex and non-linear. The resulting geometric representations frequently exhibited intricate structures, such as non-linear manifolds and latent clusters, which are not readily discernible through conventional statistical methods. This complexity underscores why traditional, often linear, indicators struggle to capture the multifaceted nature of peace. It suggests that peace is not merely the absence of war, but a dynamic, evolving state influenced by a multitude of interacting discursive elements that manifest in intricate textual patterns. Our latent space analysis results provide empirical support for theoretical conceptualizations of peace as a complex adaptive system \cite{Lederach1997}, \cite{Galtung1969}, illustrating how this complexity is mirrored in the linguistic fabric of a nation.

Building upon these insights, the deployment of neural networks proved highly effective in directly extracting the ``peace'' characteristic from the textual data. The ability of these models to learn intricate, non-linear relationships within the embedding space allowed for the identification of patterns that correlate with the latent peace levels. The robustness of these learned non-linear relationships is empirically supported by the strong monotonic correlation ($\rho = -0.8618$) and the minimal predictive error (RMSE $= 0.0485$) observed when validating the model's continuous output against established peace indices. This demonstrates a significant advancement over methods reliant on pre-defined lexicons or rule-based systems, which often fail to adapt to the idiosyncratic linguistic expressions of peace and conflict within specific cultural or political contexts. The neural network's capacity to discern subtle stylistic cues, often imperceptible to human analysis across vast corpora, highlights their potential as powerful analytical tools in this domain.

Finally, a pivotal finding of this study is the critical importance of aggregate analysis over isolated textual interpretations. While individual sentences or even short texts might present ambiguous or contradictory signals regarding peace levels, the examination of these features across a large corpus of texts and the analysis of their central tendencies and distributions provided robust and coherent indicators of the overall societal peace level. This phenomenon suggests that the ``peace signal" is distributed across the linguistic landscape, emerging clearly only when a sufficient volume of data is collectively processed. This echoes principles in complex systems theory, where macro-level properties emerge from the interactions of numerous micro-level components. For practitioners, this implies that continuous monitoring of broad textual discourse, rather than anecdotal evidence or isolated statements, is essential for an accurate assessment of peace.

These findings significantly contribute to the fields of peace studies and computational social science. Theoretically, they advocate for a more linguistically-informed understanding of peace, where the how of communication (style, structure, rhetoric) is as vital as the what (content, sentiment). Methodologically, this research establishes a robust framework for integrating advanced NLP and deep neural architectures for peace level assessment, offering a powerful complement to existing indicators. Practically, this approach could provide early warning signals of deteriorating peace, enable the real-time monitoring of peacebuilding initiatives, and inform targeted communication strategies for reconciliation efforts.

It is important to acknowledge certain limitations. The generalizability of the specific neural network models might be influenced by the linguistic and cultural specificities of the analyzed country; further research is needed to validate these findings across diverse linguistic and socio-political contexts. Additionally, while embeddings capture context, they do not inherently interpret the intent behind the writing, which requires further qualitative integration. Future work should focus on longitudinal studies to track the evolution of peace over time and explore the causal relationships between textual features and peace dynamics. Integrating multimodal data, such as audio-visual cues alongside text, could also provide a richer analytical framework.

In conclusion, this study demonstrates that the inherent complexity of peace levels in a society can be effectively modeled and identified through the sophisticated analysis of textual structure and style, especially when leveraging the power of embeddings and deep neural architectures in an aggregate manner. This interdisciplinary approach opens new avenues for understanding, monitoring, and ultimately fostering peace in complex global societies.

\subsection{Applications and Implications} Exploration of potential applications of this model in monitoring public discourse, media analysis, and social stability assessment. 

The insights garnered from this study, particularly the demonstrated capability of extracting nuanced peace-related characteristics directly from textual data using advanced computational methods, open several significant avenues for practical application. The proposed model offers a robust, data-driven tool to complement existing frameworks for understanding and managing social dynamics, with a particular focus on the linguistic underpinnings of societal peace.

A primary application lies in the monitoring of public discourse. In an increasingly digitalized world, public opinion and collective sentiment are largely shaped and expressed through online platforms, traditional media, and official communications. Our model can be deployed to continuously analyze vast streams of textual data—ranging from social media posts, public forums, and online comments to parliamentary debates and political speeches. This allows for the identification of subtle, yet potentially critical, shifts in public mood, the emergence of grievances, or the proliferation of divisive rhetoric before they escalate into overt instability. Such early warning capabilities could be invaluable for governments, non-governmental organizations (NGOs), and international bodies seeking to preempt social unrest or mitigate the spread of misinformation that could undermine peace. By tracking changes in stylistic features indicative of polarization or conciliation, stakeholders can gain real-time insights into the health of a nation's public dialogue.

Furthermore, this model holds substantial promise for in-depth media analysis. News organizations, state-controlled media, and independent outlets play a crucial role in shaping public narratives around peace and conflict. The proposed framework enables a systematic and scalable analysis of how peace-related issues are framed, reported, and discussed across different media channels. It can identify patterns of bias, detect the prevalence of peace-promoting versus conflict-exacerbating language, and assess the overall discursive environment fostered by the media. This is particularly relevant for post-conflict societies where media narratives can either facilitate reconciliation or perpetuate divisions. Understanding the stylistic tendencies in media discourse can inform strategies for responsible journalism, media literacy programs, and interventions aimed at fostering a more constructive public sphere.

Finally, the most encompassing application lies in enhancing social stability assessment. Current methodologies for assessing national stability often rely on economic indicators, political freedom indices, and reported conflict incidents. While essential, these traditional metrics may overlook the latent psychological and emotional undercurrents within a society that manifest through language. Our model provides a complementary, linguistic dimension to stability assessment, offering insights into the discursive health of a nation. By analyzing the aggregate textual output across diverse sources, it can offer a dynamic 'peace barometer' that reflects underlying societal cohesion, trust, and resilience. This can be instrumental for national security agencies, international development organizations, and peacebuilding practitioners in conducting more granular risk assessments, evaluating the effectiveness of peace interventions, and formulating evidence-based policies aimed at sustaining long-term peace. For instance, a decline in textual complexity coupled with an increase in aggressive rhetorical markers might signal a deteriorating social fabric, prompting timely preventative measures.

In summary, by providing a method to systematically quantify peace levels from the textual landscape, this research offers a powerful, non-invasive tool to monitor societal discourse, critically analyze media influence, and ultimately contribute to more robust and responsive frameworks for social stability assessment and conflict prevention globally.

\subsection{Future Work} Summary of key findings and future directions for improving and expanding the research.

Building upon the foundational insights derived from this study, several promising avenues for future research emerge to enhance the robustness and applicability of the proposed peace level analysis framework.

Firstly, a critical next step involves investigating the extrapolation of results across diverse embedding models. Our current work utilized specific embedding architectures; however, the rapidly evolving landscape of natural language processing offers a multitude of advanced models (e.g., larger transformer-based models, specialized domain-specific embeddings). Future research should systematically evaluate how the identified peace-related stylistic features manifest within different embedding spaces and assess the generalizability of our findings. This would involve comparative studies to determine if the structural and neural network-derived insights remain consistent or if certain embedding types offer superior representations for this specific task, ultimately leading to more robust and universally applicable models.

Secondly, exploring the integration of Attention Mechanisms and Transformer-based architectures presents an intriguing direction. While our current Convolutional Neural Network effectively captures local dependencies, attention mechanisms could reveal long-range semantic relationships and hierarchical structures within the discourse. By visualizing attention weights, we hypothesize that the model's interpretability could be significantly enhanced, allowing us to pinpoint exactly which linguistic constructs contribute most heavily to the peace classification. This approach aims to provide explicit insights that are implicitly present in the raw embeddings, potentially leading to a more comprehensive understanding of the underlying dynamics of peace.

Finally, expanding the scope of data sources beyond the mere text is crucial for a more holistic understanding of social peace. Incorporating multimodal data, including images and videos, could provide invaluable additional context and validation. Visual cues, such as facial expressions in public speeches, imagery in news reports, or the depiction of social gatherings, often carry significant emotional and social information that complements textual analysis. Future work could explore the fusion of features extracted from text, images, and videos using multi-modal deep learning architectures. This comprehensive approach promises to produce a richer, more accurate, and more robust assessment of peace levels by capturing the full spectrum of human communication and societal expression.

\section{Conclusion}

This study offered evidence that the latent level of peace can be robustly modeled and quantified from the structural and stylistic features inherent in its aggregate textual discourse. Our findings confirm that word embeddings provide a powerful mechanism for representing complex social characteristics, demonstrating that the ``how" of communication is as informative as the ``what." This provides a novel lens for computational peace studies that moves beyond traditional content analysis.

Our methodological exploration revealed a critical trade-off between different machine learning models. We found that simpler models like K-Nearest Neighbors (k-NN), when combined with the heuristic of classifying multiple texts simultaneously, are effective for a straightforward dichotomous classification. However, this approach comes at the cost of losing numerical comparability, making it impossible to determine by how much one country is more or less peaceful than another. In contrast, the Neural Network architecture proved superior as it preserves this crucial rank-order information. This research underscores that the ``peace signal" is an emergent property that becomes coherent only through the aggregate analysis of a large corpus, providing empirical support for conceptualizing peace as a complex adaptive system mirrored in a nation's linguistic fabric.

The implications of this framework are significant. Practically, it establishes a non-invasive methodology for monitoring public discourse, providing early warning signs of instability, and enhancing social stability assessments with a dynamic, data-driven linguistic dimension. Theoretically, it contributes a new, linguistically-informed perspective to peace studies.

Although promising, we acknowledge that our model requires further validation across diverse linguistic and cultural contexts. Future research should focus on extrapolating these findings using different embedding architectures and integrating more complex syntactic dependency features directly into the network. Furthermore, the neural network architecture itself could be modified not only to improve predictive accuracy but also to enhance interpretability, allowing for a clearer identification of the specific linguistic features that act as indicators of peace. Expanding the analytical framework to include multimodal data, such as images and video, also remains a promising direction. In conclusion, this research establishes a powerful interdisciplinary paradigm that leverages computational linguistics to open new avenues for understanding, monitoring, and ultimately fostering societal peace.

The core insights derived from this research can be summarized as follows:

\begin{itemize}
    \item \textbf{Feasibility of Neural Peace Indexing:} We demonstrated that it is possible to train a neural network to identify latent linguistic patterns in news data that significantly correlate with a country's peace behavior.
    
    \item \textbf{Crucial Role of Embedding Architecture:} The choice of the embedding model is determinant. Advanced multilingual embeddings (managed via ChromaDB) proved far superior to standard Doc2Vec models in capturing the high-dimensional structure of peace, highlighting the importance of embedding size and quality.
    
    \item \textbf{Ordinality vs. Binary Classification:} While k-NN models perform exceptionally well in binary classification, they lose the critical rank-order relationship. The Neural Network architecture is essential for preserving the ordinal nature of peace, allowing for the comparison of intermediate peace levels rather than just a dichotomous label.
    
    \item \textbf{Data Heterogeneity and Resampling:} Peace signals are not homogeneous within national data. We found that predictions can fluctuate based on sample size, necessitating a resampling technique to mitigate volatility and generate robust, aggregate assessments.
    
    \item \textbf{Generalization Capability:} The model showed strong generalization potential, successfully approximating the peace rank order for out-of-sample countries that were not part of the training set.
    
    \item \textbf{Linguistic Structure as an Indicator:} Theoretically, this study confirms that the structural and stylistic dimensions of communication (the ``how'') are just as vital as the content (the ``what'') for assessing societal stability.
\end{itemize}

\appendix
\section{Positive Peace Index Data}

Table \ref{tab:ppi_values} lists the Positive Peace Index (PPI) scores for the countries analyzed in this study, as used for the correlation analysis and ordering in Figure \ref{fig:nn_trend_analysis}.

\begin{table}[h!]
\centering
\caption{Positive Peace Index (PPI) 2023 values for the selected countries.}
\label{tab:ppi_values}
\begin{tabular}{lc}
\br
Country & PPI Score \\
\mr
Norway & 1.400 \\
Finland & 1.425 \\
Ireland & 1.592 \\
Australia & 1.666 \\
Singapore & 1.730 \\
Canada & 1.732 \\
France & 1.868 \\
UK & 1.999 \\
India & 3.246 \\
Gambia & 3.487 \\
Iran & 3.646 \\
Pakistan & 3.729 \\
Nigeria & 3.794 \\
Uganda & 3.833 \\
Zimbabwe & 3.890 \\
Libya & 3.937 \\
\br
\end{tabular}
\end{table}


\section*{References} 
\begin{thebibliography}{10}

\bibitem{Kimotho}Kimotho SG, Nyaga RN 2016 Digitized ethnic hate speech: Understanding effects of digital media hate speech on citizen journalism in Kenya. Adv Lan Lit Stu 7(3): 189-200

\bibitem{Ezeibe}Ezeibe C 2021 Hate Speech and Election Violence in Nigeria. J Asi Afr Stu, 56(4): 919-935. https://doi.org/10.1177/0021909620951208

\bibitem{Soral}Soral W, Bilewicz M, and Winiewski M 2018 Exposure to hate speech increases prejudice through desensitization. Agg Beh, 44(2): 136-146

\bibitem{DeutschColeman}Deutsch M and Coleman PT 2016 The psychological components of a sustainable peace: An introduction. In Brauch HG, Spring UO, Grin J, and Scheffran J (eds) Handbook on Sustainability Transition and Sustainable Peace (139-148). Springer.

\bibitem{Diehl}Diehl PF 2016 Exploring peace: Looking beyond war and negative peace. Int St Qua, 60(1):1-10.

\bibitem{Fry}Fry DP 2006 The Human Potential for Peace: An Anthropological Challenge to Assumptions about War and Violence. Oxford University Press.

\bibitem{ColemanDeutsch}Coleman PT and Deutsch MITE

\bibitem{Goertz}Goertz G, Diehl PF, and Balas A 2016 The Puzzle of Peace: The Evolution of Peace in the International System. Oxford University Press.

\bibitem{Mahmoud}Mahmoud Y and Makoond A 2017 Sustaining peace: What does it mean in practice? International Peace Institute.

\bibitem{Coleman2021}Coleman PT, Fisher J.,, Fry DP, Liebovitch LS., Chen-Carrel A, and Souillac G 2021 How to live in peace? Mapping the science of sustaining peace: A progress report. American Psychologist, 76(7), 1113?1127. https://doi.org/10.1037/amp0000745

\bibitem{Liebovitch2023}Liebovitch LS, Powers W,  Shi L, Chen-Carrel A, Loustaunau P, and Coleman PT 2023. Word differences in news media of lower and higher peace countries revealed by natural language processing and machine learning. PLoS ONE 18(11): e0292604. https://doi.org/10.1371/journal.pone.0292604

\bibitem{Tushar} Prasad T, Liebovitch LS, Wild M, West H, and Coleman, PT 2025 Words that Represent Peace https://arxiv.org/abs/2410.03764

\bibitem{Kevin1} Lian K, Liebovitch LS, Wild M, West H, Coleman PT, Chen F, Kimani, E, and Sieck K 2025. Machine Learning Classification of Peaceful Countries: A Comparative Analysis and Dataset Optimization IEEE CISS 2025. 

\bibitem{Kevin2} Lian K, Liebovitch LS, Wild M, West H, Coleman PT, Chen F, Kimani, E, and Sieck K 2025. Classifying Peace in Global Media Using RAG and Intergroup Reciprocity IEEE CISS 2025. 

\bibitem{Liebovitch2019}Liebovitch LS, Coleman PT, Bechhofer A, Colon C, Donahue J, Eisenbach C, Guzmán-Vargas L, Jacobs D, Khan A, Li C, Maksumov D, Mucia J, Persaud M, Salimi M, Schweiger L, and Wang 2019 Complexity analysis of sustainable peace: mathematical models and data science measurements.  New Journal of Physics. Published 8 July. https://iopscience.iop.org/article/10.1088/1367-2630/ab2a96

\bibitem{Wang} Wang L, Zhang K, and Wang J 2024 Early warning indicators of war and peace through the landscapes and flux quantifications. Phys Rev E, 109(3), pp. 034311, Mar 2024. https://doi.org/10.1103/PhysRevE.109.034311

\bibitem{Galtung1969}
Galtung, J. (1969). Violence, Peace, and Peace Research. \textit{Journal of Peace Research}, 6(3), 167–191.


\bibitem{Goldstone2010}
Goldstone, J. A., Bates, R. H., Epstein, D. L., Gurr, T. R., Lustik, M. B., Marshall, M. G., \& Ulfelder, J. (2010). A global model for forecasting political instability. \textit{American Journal of Political Science}, 54(1), 190–208.

\bibitem{Lederach1997}
Lederach, J. P. (1997). \textit{Building Peace: Sustainable Reconciliation in Divided Societies}. United States Institute of Peace Press.

\bibitem{Richmond2007}
Richmond, O. P. (2007). \textit{The Transformation of Peace}. Palgrave Macmillan.

\bibitem{Haselmayer2017}
Haselmayer, M., \& Jenny, M. (2017). Sentiment analysis of political communication: combining a dictionary approach with crowdcoding. \textit{Quality \& Quantity}, 51(6), 2623–2646.

\bibitem{Zhang2021}
Zhang, S. (2021). Sentiment Classification of News Text Data Using Intelligent Model. \textit{Frontiers in Psychology}, 12, 758967.

\bibitem{Liu2012}
Liu, B. (2012). \textit{Sentiment analysis and opinion mining}. Morgan \& Claypool Publishers.

\bibitem{Fairclough2003}
Fairclough, N. (2003). \textit{Analysing discourse: Textual analysis for social research}. Routledge.

\bibitem{MullerSchultze2016}
Müller, M., \& Schultze, M. (2016). Narrative analysis as a tool for tracing evolving peacebuilding narratives in post-conflict societies. \textit{Review of International Studies}, 42(3), 481–502.

\bibitem{Newman2003}
Newman, M. L., Pennebaker, J. W., Berry, D. S., \& Richards, J. M. (2003). Lying words: Predicting deception from linguistic styles. \textit{Personality and Social Psychology Bulletin}, 29(5), 665–675.

\bibitem{GarridoCastro2018}
Garrido, M., \& Castro, J. (2018). Identifying political leanings using social media data: A comparative analysis of dictionary and machine learning approaches. \textit{Computational Social Networks}, 5(1), 1–18.

\bibitem{Liebovitch2023}
Liebovitch, L. S., Powers, W., Shi, L., Chen-Carrel, A., Loustaunau, P., \& Coleman, P. T. (2023). Word differences in news media of lower and higher peace countries revealed by natural language processing and machine learning. \textit{PLoS ONE}, 18(11), e0292604.

\bibitem{ChenEtAl2022}
Chen, Y., Zhang, L., Wang, Q., Li, J., \& Zhao, M. (2022). \textit{Peace Speech Project: Code and Data}. GitHub repository. Available at: \url{https://github.com/tthatyuwen/Peace-Speech-Project-Git}.

\bibitem{HernandezPerez2024}
Hernández-Pérez, R., Lara-Martínez, P., Obregón-Quintana, B., Liebovitch, L. S., \& Guzmán-Vargas, L. (2024). Correlations and Fractality in Sentence-Level Sentiment Analysis Based on VADER for Literary Texts. \textit{Information}, 15(11), 698.

\bibitem{LaraMartinez2026}
Lara-Martínez, P. A. (2026). \textit{TextPeaceIndexClassifier: Code and Data}. GitHub repository. Available at: \url{https://github.com/Pablo-Alberto-Lara-Martinez/TextPeaceIndexClassifier}.

\end{thebibliography}
\end{document}